\documentclass[sigconf,nonacm]{acmart}
\usepackage{booktabs}
\usepackage{siunitx}

\AtBeginDocument{%
  }

\setcopyright{acmlicensed}
\copyrightyear{2025}
\acmYear{2025}
\acmConference[C+J '25]{Computation + Journalism Symposium 2025}{December 11--12,
  2025}{Miami, FL}

\begin{document}

\title{On-Premise AI for the Newsroom: Evaluating Small Language Models for Investigative Document Search}

\author{Nick Hagar}
\email{nicholas.hagar@northwestern.edu}
\affiliation{%
  \institution{Northwestern University}
  \city{Evanston}
  \state{IL}
  \country{USA}
}

\author{Nicholas Diakopoulos}
\email{nad@northwestern.edu}
\affiliation{%
  \institution{Northwestern University}
  \city{Evanston}
  \state{IL}
  \country{USA}
}

\author{Jeremy Gilbert}
\email{jeremy.gilbert@northwestern.edu}
\affiliation{%
  \institution{Northwestern University}
  \city{Evanston}
  \state{IL}
  \country{USA}
}

\renewcommand{\shortauthors}{Hagar et al.}

\begin{abstract}
Investigative journalists routinely confront large document collections. Large language models (LLMs) with retrieval-augmented generation (RAG) capabilities promise to accelerate the process of document discovery, but newsroom adoption remains limited due to hallucination risks, verification burden, and data privacy concerns. We present a journalist-centered approach to LLM-powered document search that prioritizes transparency and editorial control through a five-stage pipeline---corpus summarization, search planning, parallel thread execution, quality evaluation, and synthesis---using small, locally-deployable language models that preserve data security and maintain complete auditability through explicit citation chains. Evaluating three quantized models (Gemma 3 12B, Qwen 3 14B, and GPT-OSS 20B) on two corpora, we find substantial variation in reliability. All models achieved high citation validity and ran effectively on standard desktop hardware (e.g., 24 GB of memory), demonstrating feasibility for resource-constrained newsrooms. However, systematic challenges emerged, including error propagation through multi-stage synthesis and dramatic performance variation based on training data overlap with corpus content. These findings suggest that effective newsroom AI deployment requires careful model selection and system design, alongside human oversight for maintaining standards of accuracy and accountability.

\end{abstract}


\keywords{investigative journalism, semantic search, retrieval-augmented generation, large language models, hallucination detection}


\maketitle

\section{Introduction}
Investigative journalism often involves analyzing vast document collections \cite{brehmer_overview_2014}. While semantic search has emerged as a powerful tool for navigating these collections, the sheer volume of information often remains overwhelming. 

Large language models (LLMs) offer a promising solution to this analytical bottleneck. Their ability to rapidly synthesize information from multiple sources, identify patterns, and generate coherent summaries could theoretically transform weeks of document review into hours of targeted analysis \cite{cifliku_this_2025}. Retrieval-augmented generation (RAG) systems, which combine semantic search with LLM synthesis, have already demonstrated success in knowledge-intensive domains, producing fluent answers grounded in retrieved documents while maintaining citations to source material \cite{lewis_retrieval-augmented_2020}.

Yet newsroom adoption of such AI assistants remains limited \cite{cools_uses_2024}, constrained by a fundamental tension between the efficiency gains these systems promise and the verification standards journalism demands. When every published claim carries potential legal and ethical implications \cite{clary_journalists_2025}, the well-documented tendency of LLMs to hallucinate becomes a serious threat to journalistic credibility \cite{rawte_troubling_2023,openai_openai_2025}. Moreover, the proprietary and opaque nature of leading LLM services raises concerns about data privacy \cite{van_der_woude_ethics_2024} and editorial independence \cite{simon_uneasy_2022}. In this context, evaluating new LLM-driven approaches to reporting is a key concern for practice \cite{veerbeek2024using-1a6}. 

This paper presents a journalist-centered approach to LLM-powered document search that addresses these concerns through transparent, modular design. Rather than treating AI as an autonomous agent that produces final answers, we position it as a sophisticated intermediary that executes systematic search strategies while maintaining complete auditability. Our system leverages small, locally-deployable language models to preserve data security and editorial independence \cite{beggin_security_2016}, and it maintains explicit citation chains from every claim to its source documents. Through this design, we aim to capture the analytical power of AI while preserving the transparency that investigative journalism requires.

We evaluate our approach using two document collections that represent common journalistic challenges. The first mirrors the task of getting up to speed on a complex topic, requiring the system to synthesize environmental impact reports and academic papers into a coherent briefing. The second simulates an investigation, where the system acts as a research assistant to find potential story leads in a collection of legal and business documents. Our results demonstrate that small models can achieve high citation accuracy and meaningful document synthesis on standard newsroom hardware \cite{turner_lee_journalism_2024}, though with important variations in reliability that underscore the need for continued human oversight. We find that while technical capabilities like plan adherence and citation management prove robust across models, the tendency to hallucinate varies dramatically based on both model architecture and corpus characteristics, suggesting that deployment decisions must consider the investigative context.

\section{Background}
Investigative journalists often face a \textit{needle-in-the-haystack} challenge of sifting through massive troves of documents (e.g. FOIA records, leaks) on tight deadlines to find pertinent information \cite{brehmer_overview_2014}. In such large collections, keyword searches alone often fail to surface all relevant documents because journalists may not guess the exact terms hidden in the data. This limitation has motivated the adoption of \textit{semantic search}, which matches queries to results by meaning rather than literal keywords. Using now-standard techniques where documents are chunked, embedded, and indexed for fast retrieval, this capability enables journalists to discover themes, connections, and patterns that traditional search would miss \cite{cifliku_this_2025, chakraborty_scalable_2025}.

An emerging frontier in this area is the integration of generative language models with semantic search pipelines. Rather than simply returning a list of relevant documents for the journalist to read, generative systems direct an LLM to first read the retrieved documents, then produce a synthesized answer. This approach, called retrieval-augmented generation (RAG), represents an attempt to handle knowledge-intensive queries by combining the strengths of search (accuracy and grounding in up-to-date, specific information) with the strengths of LLMs (fluid natural language answers) \cite{lewis_retrieval-augmented_2020}. In theory, this yields the ``best of both worlds'': The model can cite specific source documents as evidence, which allows the user to verify claims, while retrieving relevant context reduces the model’s reliance on potentially outdated parametric memory \cite{merritt_what_2025}.

Despite the potential of RAG, newsroom adoption of such AI assistants has been cautious \cite{cools_uses_2024}. Because of their tendency to hallucinate, many newsrooms do not deem LLM outputs themselves trustworthy, requiring humans to verify information against source documents \cite{becker_policies_2025,jordaan_divining_2025}. And from a technical perspective, while LLM capabilities have increased dramatically over the past five years, hallucinations remain an issue, in some cases even increasing as models become more capable \cite{rawte_troubling_2023,openai_openai_2025}.

For newsrooms, these concerns motivate a focus on \textit{automating routine news production} and \textit{enabling faster research} while preserving transparency \cite{cifliku_this_2025,jordaan_divining_2025}. Rather than tools that act as autonomous black boxes, journalists see the largest benefit from AI systems that augment their capabilities while keeping them in the driver’s seat.

The technology industry's answer to this challenge is a new class of agentic ``deep research'' tools capable of autonomous synthesis \cite{li_reportbench_2025}. While powerful, their operation as opaque, cloud-based services makes them fundamentally unsuitable for sensitive investigative work that demands confidentiality and source protection. This points to the need for a parallel approach designed for the newsroom: a journalistic deep research system that combines sophisticated reasoning with the transparency, security, and auditable citation chains that the profession requires.

This motivates our central research question: \textbf{How can we design an LLM-powered document search assistant that amplifies journalists' investigative capabilities while preserving transparency and control?}

\section{System design}
Our system design draws on key considerations in supporting investigative reporting for newsrooms, and on established best practices and design patterns for LLM-based agents. 

\subsection{Newsroom Considerations}
\paragraph{Privacy and Security.} The document collections that investigative journalists need to search are often highly sensitive, containing material that carries legal risks \cite{clary_journalists_2025} and privacy concerns \cite{van_der_woude_ethics_2024}. As such, journalists require a high level of data security for any technical systems that will interact with these materials, ranging from restricted access to airgapped computers \cite{beggin_security_2016}. Commercial cloud services are often not sufficient for these use cases; rather, on-premise tools allow the level of control and legal protection required.

\paragraph{Resource Constraints.} Many newsrooms do not operate with extensive budgets for technical infrastructure or software engineering expertise \cite{turner_lee_journalism_2024}. As such, technical solutions must be \textit{lightweight} (i.e., not requiring specialized machines) and \textit{easily maintainable}. 

\paragraph{Transparency.} Finally, the high-stakes nature of investigative journalism means published claims must be verifiable against primary sources. In the context of an agentic system, this means that every step requires transparency into the LLM's operation, as well as direct citations to source documents---no part of the workflow can rely solely on the LLM's output without verification. 

\subsection{Agent Best Practices}
Rather than diving straight into a task or series of tasks, effective agents often first carry out a planning step---a pattern that allows the LLM to enumerate a checklist or sequence of steps, ensuring a comprehensive, high-level understanding of the goal \cite{anthropic_building_2024}. After planning, effective agents often execute searches in isolated threads, before combining them into a final answer \cite{anthropic_how_2025}. Finally, the increasing capabilities of small language models---both generally and for tool use---make them increasingly viable for agentic use cases \cite{belcak_small_2025}. 

\subsection{System Implementation}
We deploy a five-stage, on-premise pipeline designed for small, locally run LMs and transparent tool use. Each stage emits plain-text artifacts for auditability, and every claim carries an explicit citation key---generated via document chunk hashing---that resolves to the exact source passage. Full prompts, code, hyperparameters, and system outputs are available in the project repository.\footnote{\url{https://github.com/NHagar/semantic-search-assistant}}

\textbf{Stage 1--Corpus synopsis:} Summarize document headers/snippets to produce an outline of topics, recurring themes, and gaps. 
\begin{itemize}
    \item \textit{Example: For the Trump Scotland corpus, the synopsis might highlight key categories of documents like ``Scottish planning and environmental assessments'' and ``U.S. legal and court documents.''}
\end{itemize}

\textbf{Stage 2--Search planning:} Given the synopsis and a prompt specifying the required output format for search plans, propose 5--7 independent threads with objectives, sub-objectives, and candidate queries.
\begin{itemize}
    \item \textit{Example: For the AI \& environment corpus, a search thread might direct the model to ``investigate AI applications in e-waste management.''}
\end{itemize}

\textbf{Stage 3--Execution \& reporting:} Run each thread in isolation over a local semantic index; draft a thread report with citations plus open questions/next-step queries. 
\begin{itemize}
    \item \textit{Example: Reports contain claims (``For 2015, the company incurred a loss of £1.650k, with total equity at a deficit of £13,682k'') and citation keys (``[7db3cb:0]'').}
\end{itemize}

\textbf{Stage 4--Evaluation:} A lightweight judge scores relevance and coverage; only passing reports proceed to synthesis. 

\textbf{Stage 5--Synthesis:} Merge surviving reports into an executive brief, de-duplicating findings and aggregating next steps while preserving end-to-end citation chains.
\begin{itemize}
    \item \textit{Example: An executive summary for the AI \& environment corpus might contain high-level information on electricity and water usage, e-waste concerns, and any gaps in the corpus.}
\end{itemize}

\textit{Practical notes.} The system runs on a single workstation; all plans, reports, metrics, and model versions are logged for reproducibility. Depending on the model and corpus, a typical end-to-end run of this system took 2–3 hours.  Implementation details (indexing, chunking, embedding model, vector store) are documented in the repository.

\section{Evaluation}
We evaluate our system design on two tasks: A reference task (i.e., getting up to speed on a high-level collection of documents) and an investigation task (i.e., highlighting potential leads in a document corpus). 

For the reference task, we collect 64 news articles, academic articles, and technical reports cited in the Wikipedia article on the environmental impact of AI\footnote{https://en.wikipedia.org/wiki/Environmental\_impact\_of\_artificial\_intelligence}. For the investigation task, we leverage the ``Trump Scotland'' corpus on Aleph\footnote{https://aleph.occrp.org/datasets/2705}, a database maintained by the Organized Crime and Corruption Reporting Project. Compiled by Adam Davidson, this corpus contains documents---legal filings, building designs, permits, and correspondence---pertaining to Donald Trump's golf course in Aberdeen, Scotland. We extract 292 documents from this corpus. 

We test our system with three small language models, chosen to provide a balance of model capability and efficiency: Gemma 3 12B\footnote{Due to memory constraints on our hardware, the Gemma 3 12B model could not process the high token count of the ``Trump Scotland'' corpus required for the initial synopsis stage. This condition is therefore omitted from our results.} \cite{team_gemma_2025}, Qwen 3 14B \cite{qwen3technicalreport}, and GPT-OSS 20B \cite{openai_gpt-oss-120b_2025}. All models are quantized to 4-bit precision. All evaluation runs are executed on a MacBook Air with an M3 chipset and 24 GB of unified memory, using LM Studio. 

\subsection{Protocol}
We run the system with identical model configurations for each dataset. For each run, we collect all thread reports (6--7 reports per run) and the final synthesized report. We then conduct claim-level annotation on each thread report. We segment outputs into atomic claims and, for each claim, record support status, citation validity, and (if applicable) hallucination severity.

\subsection{Metrics}
We report four metrics, averaged across all thread reports, to capture factuality, hallucination risk, citation validity, and process reliability (exact calculation details in the repository).

\paragraph{Claim support rate.} Share of model claims that are directly supported by at least one cited passage from the corpus.

\paragraph{Hallucination severity index (HSI).} Counts hallucinated claims and weights them by severity---minor (1), major (2), critical (3)---then averages across reports; lower is better \cite{rawte_troubling_2023}.

\paragraph{Invalid citation rate.} Fraction of citations that do not correspond to actual document chunks.

\paragraph{Plan adherence.} Fraction of planned sub-objectives that were either satisfied with supported evidence or explicitly concluded as unsupported after documented search attempts.

\subsection{Final Reports}
Because the final reports for each search are generated via a synthesis of the thread reports, we do not reexamine errors that were already present in the thread reports. Instead, we report instances where the final report introduced additional unsupported claims, invalid citations, or hallucinations into the synthesized material. 

\section{Results}

Table \ref{tab:model_performance} summarizes the performance of our three tested models across both corpora. The results reveal substantial variation in model reliability: Qwen 3 14B demonstrated the most consistent performance across all metrics, while GPT-OSS 20B showed more variable outcomes depending on the corpus and task complexity.

\begin{table}[htbp]
  \centering
  \small 
  \caption{Model performance (search thread-level averages) across corpora. Qwen 3 14B reliably produces valid citations and the fewest hallucinations.}
  \label{tab:model_performance}
  \begin{tabular}{@{}ll S[table-format=1.2] S[table-format=2.2] S[table-format=1.2] S[table-format=1.2]@{}}
    \toprule
    Model & Corpus & {Claim} & {Halluc.} & {Invalid} & {Plan} \\
          &        & {Support} & {Index}   & {Citation}  & {Adherence} \\
    \midrule
    Qwen    & AI \& env. & 0.95 & 1.83  & 0.00 & 1.00 \\
    Gemma   & AI \& env. & 1.00 & 8.57  & 0.10 & 0.79 \\
    GPT-OSS & AI \& env. & 0.92 & 22.50 & 0.00 & 1.00 \\
    \midrule
    Qwen    & Trump Scot. & 0.98 & 1.00  & 0.00 & 1.00 \\
    Gemma   & Trump Scot. & {--} & {--}    & {--}   & {--}   \\
    GPT-OSS & Trump Scot. & 1.00 & 1.50  & 0.00 & 1.00 \\
    \bottomrule
  \end{tabular}
\end{table}

\subsection{Instruction Following and System Reliability}
With the exception of Gemma 3 12B, the tested models demonstrated strong capabilities in following multi-stage pipeline instructions. Both Qwen 3 14B and GPT-OSS 20B achieved perfect plan adherence scores, addressing all specified sub-objectives within their assigned search threads. This high adherence rate indicates that these models can reliably execute planned search strategies. The models also demonstrated robust citation management, accurately replicating citation keys that corresponded to actual document chunks retrieved from the vector index. This technical reliability provides a crucial foundation for journalistic use cases. 

\subsection{Validity of Generated Reports and Hallucination Patterns}
The validity of generated reports varied considerably across models and corpora, revealing important patterns in how different architectures handle document-grounded reasoning tasks. Across all tested conditions, the vast majority of claims aligned correctly with their provided citations, indicating that when models cite sources, they generally do so accurately. However, variation emerged in the frequency and severity of hallucinations.

Gemma 3 12B proved the most problematic, generating substantial hallucinations with a severity index of 8.57 on the AI \& environment corpus. The model's poor performance extended beyond hallucinations to plan adherence, managing only 79\% completion of its planned sub-objectives.

Qwen 3 14B emerged as the most reliable performer, producing the equivalent of 1-2 minor hallucinations per report across both test corpora. Its consistency across different document types and investigation contexts suggests robust grounding capabilities that align well with journalistic verification requirements.

GPT-OSS 20B showed marked corpus sensitivity, performing well on the Trump Scotland documents (HSI: 1.50) but poorly on the AI \& environment collection (HSI: 22.50). This difference may be linked to the model's tendency to incorporate information from its training data. On the AI \& environment corpus---a topic likely well-represented in the model's training---GPT-OSS frequently referenced documents and facts that it had not actually encountered during the search process. Notably, this pattern did not emerge with the Trump Scotland corpus, possibly because this specialized legal and business documentation represents a more niche domain with less representation in model training data.

\subsection{Final Report Synthesis Challenges}
Our analysis of the final synthesized reports revealed systematic issues with the multi-stage aggregation process. The system lacks any correction mechanism, meaning hallucinated or incorrect information from individual thread reports propagates unchanged into the final output. This error persistence represents a fundamental challenge for multi-stage agentic systems in high-stakes domains like journalism. We also observed one concerning instance where Qwen 3 14B, despite its generally strong performance at the thread level, introduced two severe hallucinations and one invalid citation during the synthesis stage of the Trump Scotland corpus analysis. This finding suggests that even reliable models can introduce errors during the reasoning required to merge information threads. 

\subsection{Qualitative Observations and System Limitations}
Several qualitative patterns emerged that illuminate both capabilities and constraints of the current system design. Rather than producing entirely fabricated reports when encountering sparse search results, both Qwen 3 14B and GPT-OSS 20B demonstrated appropriate uncertainty handling. They reliably flagged instances where no relevant documents were retrieved for specific search queries. GPT-OSS proved especially transparent in this regard, producing detailed logs of attempted searches when encountering dead ends on investigative threads.

\subsubsection{Corpus Composition and Breadth Effects}
The relationship between corpus composition and search performance proved complex and context-dependent. The Trump Scotland corpus presented particular challenges due to its heterogeneous document types, ranging from architectural renderings to legal filings. In our current configuration, search agents struggled with this diversity, focusing predominantly on the corpus's primary themes (building plans and permitting processes) while overlooking potentially valuable auxiliary material.

The AI \& environment corpus maintained much tighter topical focus, yet produced less reliable results from GPT-OSS, suggesting that corpus breadth alone does not determine efficacy. The interaction between model architecture, training data overlap, and document diversity requires further investigation to optimize system performance across different investigative contexts.

\subsubsection{Document Processing and Technical Infrastructure}
Document preprocessing quality emerged as a critical factor in search effectiveness. OCR errors and malformed text occasionally confounded the models' ability to locate relevant chunks, underscoring the importance of high-quality document ingestion pipelines. However, the models also demonstrated surprising robustness, successfully extracting precise figures from malformed tabular data and correcting minor OCR errors during report generation. This suggests that while perfect document processing is ideal, the underlying system can generate useful results even with imperfect input data.

\section{Discussion}
Our evaluation demonstrates that small language models can serve as capable research assistants for investigative journalism, though with important caveats that illuminate broader questions about human--AI collaboration in high-stakes domains. While models like Qwen 3 14B achieved near-perfect citation accuracy and plan adherence, the substantial variation in hallucination rates and the systematic error propagation during synthesis underscore that effective deployment requires careful model selection, transparent system design, and sustained human oversight.

These findings suggest the path forward is not a choice between human expertise and AI, but a need for systems that amplify journalistic capabilities while preserving verification standards. \cite{diakopoulos_algorithmic_2017}. Rather than attempting to make model reasoning interpretable---a potentially impossible task \cite{shojaee2025illusionthinkingunderstandingstrengths}---our approach maintains a clear chain of evidence from query to citation to source document. Our multi-stage design also demonstrates that pursuing transparency does not require abandoning sophisticated analytical capabilities. Through careful prompt engineering and modular architecture, the system provides systematic document exploration that would be prohibitively time-consuming for human researchers, while ensuring that every claim remains traceable to source material.

As with many newsroom technologies, deploying LLMs is an editorial, not just technical, decision \cite{hagar_optimizing_2019}. Our results show that model selection reflects fundamental choices about verification standards. Editors must consider how a model's training might bias its performance on certain topics. For example, a model that excels on niche legal documents might prove unreliable on topics well-represented in its training data, where it may confuse retrieved evidence with its own parametric knowledge.

\subsection{Practical Implications}
Our findings address two critical barriers to newsroom AI adoption: cost and data security. The ability to run effective semantic search and synthesis pipelines on a single workstation with 24 GB of memory means that even resource-constrained newsrooms can experiment with AI assistance \cite{kleppmann_local-first_2019}. This finding is particularly important for investigative teams working with sensitive documents that cannot be uploaded to commercial AI services.

However, successful implementation requires recognizing these systems as extensions of journalistic capability rather than replacements for editorial judgment. Our results suggest that the most productive deployment approach treats AI as a sophisticated research assistant that can surface relevant material and salient patterns, while leaving editorial decisions about significance, newsworthiness, and story direction firmly in human hands.

\subsection{Limitations}
Our evaluation prioritized functional accuracy and citation validity over editorial judgment. Future work should incorporate newsworthiness assessments. Our test corpora, while drawn from real investigative contexts, may differ from typical document dumps in ways that affect system performance. In particular, this system design focuses solely on text-based research. Future work should explore evaluation approaches for multimodal corpora. Finally, the probabilistic nature of language model inference means that our quantitative results may not reflect consistent real-world performance. Small models, in particular, can show significant run-to-run variation that our limited evaluation runs may not have captured. Newsrooms considering deployment need extensive testing on their specific document types and use cases.

\subsection{Conclusion}
The integration of AI assistance into investigative journalism requires balancing sophisticated analytical capabilities with the transparency and accountability that credible reporting demands. Our findings demonstrate this balance is achievable through careful system design that prioritizes auditability, treats model selection as an editorial decision, and positions AI as an extension of rather than replacement for journalistic expertise. While significant technical and methodological challenges remain, the demonstrated capability of local, resource-efficient models suggests that newsrooms can experiment with AI-assisted investigation without compromising their core values of verification, independence, and editorial control.

\begin{acks}
This work was conducted with support from the Knight Foundation.
\end{acks}

\bibliographystyle{ACM-Reference-Format}
\bibliography{sample-base}


\end{document}